\documentclass[12pt]{iopart}

\usepackage{amssymb}
\usepackage[dvips]{graphicx}
\usepackage{colordvi}
\usepackage{times}

\newcommand{\un}[1]{\underline{#1}}

\newcommand{\bc}{\begin{center}}
\newcommand{\ec}{\end{center}}
\newcommand{\be}{\begin{equation}}
\newcommand{\ee}{\end{equation}}
\newcommand{\bna}{\begin{eqnarray}}
\newcommand{\ena}{\end{eqnarray}}

\newcommand{\mpaa}{\begin{minipage}[t]{6cm}}
\newcommand{\mpea}{\end{minipage}}
\newcommand{\mpab}{\begin{minipage}[t]{8cm}}
\newcommand{\mpeb}{\end{minipage}}
\newcommand{\mpac}{\begin{minipage}[t]{13cm}}
\newcommand{\mpec}{\end{minipage}}
\newcommand{\mpad}{\begin{minipage}[t]{13cm}}
\newcommand{\mped}{\end{minipage}}
\newcommand{\mpae}{\begin{minipage}[t]{13cm}}
\newcommand{\mpee}{\end{minipage}}
\newcommand{\mpaf}{\begin{minipage}[t]{6cm}}
\newcommand{\mpef}{\end{minipage}}

\newcommand{\za}{\alpha}
\newcommand{\zb}{\beta}
\newcommand{\zd}{\delta}
\newcommand{\ze}{\epsilon}

\newcommand{\zL}{\Lambda}

\newcommand{\zR}{I\hskip-3.4pt R}
\newcommand{\zw}{\omega}
\newcommand{\zW}{\Omega}

\newcommand{\noi}{\noindent}

\newcommand {\bdm} {\begin{displaymath}}
\newcommand {\edm} {\end{displaymath}}

\usepackage{color}
\definecolor{darkblue}{rgb}{0,0,0.6}
\definecolor{darkred}{rgb}{0.7,0,0}
\definecolor{darkgreen}{rgb}{0,0.7,0}
\usepackage{amsfonts}

\newcommand{\bea}{\begin{eqnarray}}
\newcommand{\eea}{\end{eqnarray}}

\begin{document}

\title{Steady state fluctuation relation and time-reversibility
for non-smooth chaotic maps}

\author{Matteo Colangeli$^1$, Rainer Klages$^2$, Paolo De Gregorio$^1$,
Lamberto Rondoni$^1$}
\address{$^1$Dipartimento di Matematica, Politecnico di Torino, Corso Duca degli Abruzzi 24, 10129 Torino, Italy\\
$^2$ Queen Mary University of London, School of Mathematical Sciences,\\
Mile End Road, London E1 4NS, UK}

\ead{colangeli@calvino.polito.it}

\begin{abstract}
Steady state fluctuation relations for dynamical systems are commonly
derived under the assumption of some form of time-reversibility and of
chaos. There are, however, cases in which they are observed to hold
even if the usual notion of time reversal invariance is violated,
e.g.\ for local fluctuations of Navier-Stokes systems. Here we
construct and study analytically a simple non-smooth map in which the
standard steady state fluctuation relation is valid, although the
model violates the Anosov property of chaotic dynamical
systems. Particularly, the time reversal operation is performed by a
discontinuous involution, and the invariant measure is also
discontinuous along the unstable manifolds. This further indicates
that the validity of fluctuation relations for dynamical systems does
not rely on particularly elaborate conditions, usually violated by
systems of interest in physics. Indeed, even an irreversible map
is proved to verify the steady state fluctuation relation.
\end{abstract}

\maketitle

\section{Introduction}

One of the central aims of nonequilibrium statistical physics is to
find a unifying principle in the description of nonequilibrium
phenomena. Nonequilibrium fluctuations are expected to play a major
role in this endeavor, since they are ubiquitous, are observable in
small as well as in large systems, and a theory about them is
gradually unfolding, cf.\ Refs.\
\cite{ESreview,Ruelle,RonMejia,RKbook,Bettolo,ReyBellet} for recent
reviews. A number of works have been devoted to the derivation and
test of fluctuation relations (FRs), of different nature
\cite{ECM,Gall,ES94,axiomc,LeboSpohn,Gawedzki,SearlRonEvans,CheKla}. It is
commonly believed that, although nonequilibrium phenomena concern a
broad spectrum of seemingly unrelated problems, such as hydrodynamics
and turbulence, biology, atmospheric physics, granular matter,
nanotechnology, gravitational waves detection, etc.\
\cite{Bettolo,JR10,Gonn,Rare}, the theory underlying FRs rests on
deeper grounds, common to the different fields of application. This
view is supported by the finding that deterministic dynamics and
stochastic processes of appropriate form obey apparently analogous FRs
\cite{Bettolo,ReyBellet,LeboSpohn,Gawedzki}, and by the fact that
tests of these FRs on systems which do not satisfy all the
requirements of the corresponding proofs typically confirm their
validity.  Various works have been devoted to identify the minimal
mathematical ingredients as well as the physical mechanisms underlying
the validity of FRs \cite{MorrRonAxC,SearlRonEvans,MejRo,ReyBellet}.
This way, the different nature of some of these, apparently identical
but different, FRs has been clarified to a good extent
\cite{SearlRonEvans,RonMejia,Gawedzki,ReyBellet}. However,
analytically tractable examples are needed to clearly delimit the
range of validity of FRs, and to further clarify their meaning.

In this paper, the assumptions of time reversal invariance and of
smoothness properties, required by certain derivations of FRs for
deterministic dynamical systems, are investigated by
means of simple models that are amenable to detailed mathematical
analysis.
In particular, we consider the steady state FR for the observable
known as the phase space contraction rate $\Lambda$, which we call the
$\Lambda$-FR, for dissipative and reversible dynamical systems, in
cases in which $\Lambda$ equals the so-called dissipation function
$\zW$ \cite{ESreview}, and the $\Lambda$-FR then equals the steady
state $\zW$-FR \cite{SearlRonEvans}. As will be shown below, the phase
variables $\Lambda$ and $\zW$ coincide provided that the probability density entering the definition of $\zW$ is taken uniform, as in the case of the equilibrium density for the baker map \cite{Dorfman}.
Both the $\Lambda$-FR and the $\zW$-FR rest on dynamical assumptions:
While the steady state $\zW$-FR has been proven to hold under the
quite mild condition of decay of correlations with respect to the
initial (absolutely continuous, with respect to the Lebesgue measure)
phase space distribution \cite{SearlRonEvans}, the $\Lambda$-FR has
been proven for a special class of smooth, hyperbolic (Anosov)
dynamical systems \cite{ECM,Gall}, whose natural measure is an SRB
measure.
Indeed, there are almost no systems of physical interest that strictly
obey such conditions. However, in a similar fashion, there are almost
no systems of physical interest satisfying the Ergodic Hypothesis, and
yet this hypothesis is commonly adopted and leads to correct
predictions. Analogously to the ergodic condition, one may thus
interpret the Anosov assumption as a practical tool to infer the
physical properties of nonequilibrium systems.  Nevertheless, it is
important to investigate which aspects of the derivation of the
$\Lambda$-FR are not essential to its validity. Along these lines one
notices that the $\Lambda$-FR seems to inherently rely on a rigid
notion of time reversibility, which, however, is not always satisfied
\cite{Porta}, and on the smoothness of the natural measure along the
unstable directions, which is also problematic. On the other hand, the validity of the $\zW$-FR has
never been explicitly checked on an exactly solvable model.

By considering a fundamental class of chaotic dynamical systems, known
as baker maps, we want to assess the relevance of the Anosov
assumption and of time reversibility for the validity of the
$\Lambda$-FR, in cases in which it coincides with the $\zW$-FR. Also,
by assessing the validity of these FRs while violating standard
assumptions, we probe and extend their range of validity.  The maps we
consider are appealing, since they are among the very few dynamical
systems which can be analytically investigated in full detail. For
this reason, baker maps have often been used as paradigmatic models of
systems that enjoy nonequilibrium steady states (NESS)
\cite{Dorfman,PGbook,RKbook,Voll02,RTV00}.

Our main results are summarized as follows: The assumed sufficient
conditions of the standard derivation of the $\Lambda$-FR, i.e.\
smooth time reversal operator and the Anosov property, are not
necessary. Indeed, the $\Lambda$-FR is verified in maps whose
invariant measure and time reversal involution are discontinuous along
the unstable direction. This result is connected with the fact that
$\Lambda$ equals $\zW$, and that the $\zW$-FR is known to be quite a
generic property of reversible dynamics.  The Anosov condition allows
the natural measure to be approximated in terms of unstable periodic
orbits, which constitutes a convenient tool in low dimensional dynamics
and even in some high dimensional cases \cite{MR94,Ruelle,Ladek,JR10}.
This approximation may hold even if the Anosov condition is not
strictly verified, because periodic orbits enjoy particular symmetries
which other trajectories do not \cite{ChaosBook}. However, if the
Anosov condition is violated, one must check case by case whether the
unstable periodic orbit expansion may be trusted.  We will also face
this issue, showing that in some case the unstable periodic orbit
expansion becomes problematic, hence a different approach must be
developed.  In particular, we will profit from a separation of the
full phase space into two regions, within each of which the invariant
measure as well as the time reversal operator are
smooth. Nevertheless, the full system is ergodic: the two regions are
not separately invariant, and any typical trajectory densely explores
both making the discontinuities relevant, e.g.\ for the role of
periodic orbits.

\section{Time-reversibility for maps}
\label{sec:sec1}

In this Section we review the concept of time reversibility for time discrete deterministic
evolutions.
In order to remain close to the notion of (microscopic) time
reversibility of interest to physics, one usually calls time reversal
invariant the maps whose phase space dynamics obeys a given
symmetry.
In particular, one commonly calls {\em reversible}
a dynamical system if there exists an {\em involution} in phase space,
which anticommutes with the evolution operator \cite{RQ,Ruelle}.

In practice, consider a mapping $M : \mathcal{U} \to \mathcal{U}$ of
the phase space $\mathcal{U}\subset\mathbb{R}^d\:,\:d\in\mathbb{N}$,
which evolves points according to the deterministic rule
\be
\un{x}_{n+1}=M(\un{x}_{n})\quad ,
\ee
where $n$ is the discrete time.
The set of points $\{\un{x}_{1},\un{x}_{2},\un{x}_{3},...\}$, obtained
by repeated application of the map $M$, constitutes the discrete
analogue of a phase space trajectory of a continuous time dynamical
system and, indeed, each $\un{x}_{n}$ could be interpreted as a
snapshot of the states visited by a continuously evolving system.  If
$M$ admits an inverse, $M^{-1}$, which evolves the states backward in
time, like rewinding a movie, with inverted dynamics
$\un{x}_n=M^{-1}(\un{x}_{n+1})$, $M$ is called
\textit{reversible} if there exists a transformation
$G$ of the phase space that obeys the relation
\be
G M G =M^{-1}\: ,\: G G=I \quad ,  \label{reversible}
\ee
where $I$ is the identity mapping.

This is not the only possible notion of reversibility; there exists a
variety of weaker as well stronger properties \cite{RQ,Lamb}, which
may be thought of as abstract counterparts of the time reversibility
of the dynamics of the microscopic constituents of matter.  For every
$\un{x} \in \mathcal{U}$, the symmetry property Eq.(\ref{reversible})
obviously implies
\be
G M G M (\un{x}) = \un{x} ~.
\label{invol}
\ee
If $M$ is a diffeomorphism, as often assumed \cite{Ruelle}, Eq.(\ref{invol}) can be
differentiated to obtain
\be
D G(M G M (\un{x}))D M(G M (\un{x}))D G(M (\un{x}))D M(\un{x})=I \quad ,
\ee
where $D M(\un{x})$ denotes the Jacobian matrix of $M$ evaluated at
the point $\un{x}$ of the phase space, and similarly
$DG(\un{x})$. Using the relations $[DM]^{-1}(\un{x})=D
M^{-1}(M(\un{x}))$ and $[DG]^{-1}(\un{x})=D G(G(\un{x}))$ leads to
\bea
&&D M(G M (\un{x}))D G(M (\un{x}))D M(\un{x}) = D G(M G M G (\un{x})) \nonumber\\
&& \nonumber \\
&&D G(M (\un{x}))D M(\un{x}) = D M^{-1}(M G M (\un{x}))D G(M G M G (\un{x}))\nonumber\\
&& \nonumber \\
&&D M(\un{x}) = D G(G M (\un{x})) D M^{-1}(M G M (\un{x}))D G(M G M G(\un{x}))\quad ,
\eea
which, together with (\ref{reversible}), yields
\be
D M(\un{x})=D G(G M (\un{x})) D M^{-1}(G(\un{x})) D G(\un{x}) ~.
\label{first}
\ee
Moreover, computing the determinant of the matrices in
Eq.(\ref{first}) we obtain
\be
J_M(G M (\un{x})) J_M(\un{x}) \frac{J_G(M (\un{x}))}{J_G(\un{x})}=1 \label{relation}
\ee
where $J_M(\un{x})=|\det DM(\un{x})|$ and $J_G(\un{x})=|\det DG(\un{x})|$
stand for the local Jacobian determinants computed at $\un{x}$.
Because the involution $G$ is unitary and $J_G(\un{x})=1$ for every $\un{x}$,
by definition, Eq.(\ref{relation}) can be simplified to obtain
\be
J_{M}(\un{x})=J_M^{-1}(G M (\un{x}))
\label{general}
\ee
for all $\un{x}$ in the phase space. This equation provides
a key ingredient for the derivation of fluctuation relations in dynamical systems
\cite{Gall,Gall2,Ruelle}, as we will also see later on for our
examples.

Reversible dissipative systems have been discussed extensively in
connection with so-called thermostatting algorithms, both for time
continuous \cite{Hoover,Evans,RKbook} and time discrete
\cite{VTB,Voll02,GiDo99} dynamics. Special attention has been paid in these systems to the time
average of the phase space contraction rate $\zL(\un{x})=-\ln
J_M(\un{x})$, which is an indicator of the dissipation rate. The other
indicator recently used in connection with FRs is the dissipation
function which, in our context, takes the form
\cite{SearlRonEvans,RonMejia}
\begin{equation} \label{omegat}
\zW(\un{x}) := \log \frac{\rho(\un{x})}{\rho(GM\un{x})}+ \zL(\un{x})
\end{equation}
for a given phase space probability density $\rho$. Obviously, $\zW$ takes
different forms depending on $\rho$, and one has $\zL=\zW$ if $\rho$
is uniform in the phase space, which will be our case. Hence, in the following we only use $\zL$ for
simplicity.

On a trajectory segment of duration $n$ steps,
starting at initial condition $\un{x}_{0}$, the time average of $\zL$ is defined by
\be
\overline{\zL}_{n}(\un{x}_{0}) - \frac{1}{n}\sum_{k=0}^{n-1}\ln J_M (M^{k}(\un{x}_{0})) \label{sigmadir}\quad .
\ee
Given this trajectory segment, let us call {\em reversed
trajectory segment} the segment of duration $n$ and initial condition
$G M^{n}(\un{x}_{0})=M^{-n}G(\un{x}_{0})$, cf.\
Eq.(\ref{reversible}). Its average phase space contraction rate may
be written as
\bea
\overline{\zL}_{n}(G M^{n}(\un{x}_{0}))&=&- \frac{1}{n}\sum_{k=0}^{n-1}\ln J_M (M^{k}G M^{n}(\un{x}_{0}))\nonumber\\
&=& - \frac{1}{n}\sum_{k=0}^{n-1}\ln J_M (G M^{-k+n}(\un{x}_{0})) \nonumber\\
&=& \frac{1}{n}\sum_{k=0}^{n-1}\ln J_M (M^{-k+n-1}(\un{x}_{0}))\label{sigmainv}
\eea
in which the last equality follows from Eq.(\ref{general}) if the
dynamics is time reversal invariant. We have thus shown that the phase
space contraction rates of reverse trajectories take opposite values,
\be
\overline{\zL}_{n}(G M^{n}(\un{x}_{0}))=-\overline{\zL}_{n}(\un{x}_{0}) \quad ,
\label{property}
\ee
in time-reversible dissipative systems. It is interesting to note
that, in discrete time, the initial condition of the reverse
trajectory is constructed by applying the reversal operator $G$ to a
point, $M^n(\un{x}_{0})$, which is not part of the forward trajectory
segment, but is reached one time step after the last point of the
original segment. This equation is at the heart of the proof of
steady state FRs for reversible dynamical systems.

\section{The $\Lambda$-FR}
\label{sec:sec2}

The steady state $\Lambda$-FR was first obtained by Evans, Cohen and
Morriss \cite{ECM} for a Gaussian ergostatted (i.e.\ constant energy
\cite{Hoover}) particle system, whose
entropy production rate is proportional to the phase space contraction
rate. It was then rigorously shown to be characteristic of the phase
space contraction rate of time reversal invariant, dissipative,
transitive Anosov systems by Gallavotti and Cohen \cite{Gall}.

This relation may be expressed as follows. Consider the dimensionless
phase space contraction rate, averaged over a trajectory segment of
duration $n$, with middle point $\un{x}$, in the phase space $\mathcal{U}$,
\begin{equation}
e_n(\un{x}) \frac{1}{n \langle \zL \rangle} \sum_{k=-n/2}^{n/2-1} \zL(M^k (\un{x}))
= \frac{1}{\langle \zL \rangle} ~ \overline{\zL}_n(M^{-n/2}(\un{x}))\quad ,
\label{p}
\end{equation}
where, without loss of generality, $n$ is even and
$$
\langle \zL \rangle = \int_\mathcal{M} \zL(\un{x}) ~ \mu({\rm d} \un{x})
$$
is the nonequilibrium steady state phase space average of $\zL$,
computed with respect to the natural measure $\mu$ on $\mathcal{U}$,
i.e.\ the $M$-invariant measure characterizing the time statistics of
trajectories typical with respect to the Lebesgue measure.  Then the
Fluctuation Theorem may be stated as follows \cite{Gall,Ruelle}:

\vskip 5pt
\noi
{\bf Gallavotti-Cohen Fluctuation Theorem.} {\it Let  $M$ be a $C^{1+\za}$,
$\za>0$, reversible Anosov diffeomorphism of the compact connected manifold
$\mathcal{U}$, with an involution $G$ and a $G$-invariant Riemann metric.
Let $\mu$ be the corresponding SRB measure, and assume that $\langle \zL \rangle > 0$
with respect to $\mu$. Then there exists $p^* > 0$ such that
\begin{equation}
p - \zd \le \lim_{n \to \infty} \frac{1}{n \langle \zL \rangle} \log
\frac{\mu(\{ x : e_n(x) \in (p-\zd,p+\zd)\})}
{\mu(\{ x : e_n(x) \in (-p-\zd,-p+\zd)\})} \le p+\zd
\label{prethm}
\end{equation}
if $|p| < p^*$ and $\zd > 0$.}
\vskip 5pt

\noi

Eq.(\ref{prethm}), usually considered for an arbitrarily small
$\zd$ and by specifically dealing with the phase space contraction
rate as an observable, refers to what we denoted as the $\zL$-FR in
the introduction. According to this terminology, one may say that the
Gallavotti-Cohen Fluctuation Theorem proves the $\zL$-FR under
specific conditions. This theorem is a rather sophisticated result,
obtained by heavily relying on properties of Anosov diffeomorphisms,
hence, in principle, it is hardly generic (see also \cite{Gall2}). For
instance, Ruelle's derivation
\cite{Ruelle} makes use of Bowen's shadowing property, topologically mixing
specifications, properties of sums for H\"older continuous functions,
expansiveness of the dynamics, continuity of the tangent bundle splitting,
the unstable periodic orbit expansion of $\mu$, and large deviations results
for one dimensional systems with short range interactions.  In these
derivations, time reversibility and transitivity are necessary to
ensure that the denominator of the fraction in the $\zL$-FR does not
vanish when the numerator does not, while the smoothness of the
invariant measure along the unstable directions, which allows the
periodic orbit expansion, is included in the SRB property of
$\mu$. Recently, Porta has shown for perturbed cat maps
that the $\zL$-FR requires the existence of a smooth
involution representing the time reversal operator.

Experimental and numerical verifications of relations looking like
Eq.(\ref{prethm}), for observables of interest in physics, have been
obtained for systems which may hardly be considered Anosov
\cite{ESreview,RonMejia,Cil10}. Therefore, especially in view of the
fact that the observable of interest is not $\zL$, except in very
special situations, various studies have argued that strong dynamical
properties, such as those required by the standard proof of the
fluctuation theorem for $\zL$, should not be strictly
\textit{necessary} \cite{ESR,SearlRonEvans,RonMejia,Bettolo}.  Indeed,
according to these references, time reversibility seems to be the
fundamental ingredient for fluctuation relations of the physically
interesting dissipation, since a minimum degree of chaos, such that
correlations do not persist in time, can be
taken for granted in most
particle systems.\footnote{Of course, one may expect exceptions to this rule in cases where randomness in the
dynamics is somewhat suppressed
\cite{SearlRonEvans,RonMejia}.}

Here we will proceed to show that properties implied by the Anosov
condition, like the smoothness of the natural measure along the
unstable directions, are violated in some simple models while the
$\zL$-FR still holds.

\section{The $\Lambda$-FR for a simple dissipative baker map}
\label{sec:sec3}

Research on chaos and transport has strongly benefitted from the study
of simple dynamical systems such as baker maps
\cite{Dorfman,PGbook,Voll02,RKbook}. These paradigmatic models provide
the big advantage that they can still be solved analytically, because
they are piecewise linear, yet they exhibit non-trivial dynamics which
is chaotic in the sense of displaying positive Lyapunov
exponents. There are two fundamentally different ways to generate
nonequilibrium steady states for such systems \cite{RKbook}, namely by
considering area preserving, `Hamiltonian-like' maps under suitable
nonequilibrium boundary conditions \cite{PGbook,GaKl98,GiDo99}, or by
including dissipation such that $\langle \zL \rangle > 0$, as required
by the $\zL$-FR \cite{Voll02,VTB,Dorfman}. Within the framework of
the former approach, FRs for baker maps have been derived in
Refs.~\cite{Voll02,RTV00}. Here we follow the latter
approach by endowing the map with a bias, which can be represented by
a suitable asymmetry in the evolution equation. This bias may mimic an
external field acting on the particles of a given physical system by
generating a current $\Psi$. One should further require the map to be
area contracting (expanding) in the direction parallel (opposite) to
the bias, which is the situation in standard thermostatted particle
systems \cite{VTB,Voll02}.

We now discuss the proof of the $\Lambda$-FR for maps of this
type. The probably most simple model is described in
Refs.~\cite{TGD98,GiDo99,Dorfman}. Here we give, in a different
fashion than in the book Ref.~\cite{Dorfman}, the proof of the
$\zL$-FR for this system by including the one sketched in this
book. This sets the scene for a slightly more complicated model, which
we will analyze in the following section. The calculations that follow
allow us in particular to investigate the applicability of the
unstable periodic orbit expansion for cases in which the smoothness
conditions that guarantee their applicability are violated, but to a
different extent in the different models.


Let $\mathcal{U} = [0,1] \times [0,1]$ be the phase space, and
consider the evolution equation

\be
\left(\begin{array}{c}
  x_{n+1} \\
  y_{n+1}
\end{array}\right)
=M
\left(\begin{array}{c}
  x_{n} \\
  y_{n}
  \end{array}\right)
  =\left\{
     \begin{array}{ll}
       \left(\begin{array}{c}
                   x_{n}/l \\
                   r y_{n}
                 \end{array}\right), & \hbox{for $0\leq x \leq l$;} \\
       \left(\begin{array}{c}
                   (x_{n}-l)/r \\
                   r+l y_{n}
                 \end{array}\right), & \hbox{for $l\leq x \leq 1$.}
     \end{array} \right. \label{Map1} \quad .
\ee

\begin{figure}[tbh]
\begin{center}
  \includegraphics[width=9cm, width=0.80\textwidth]{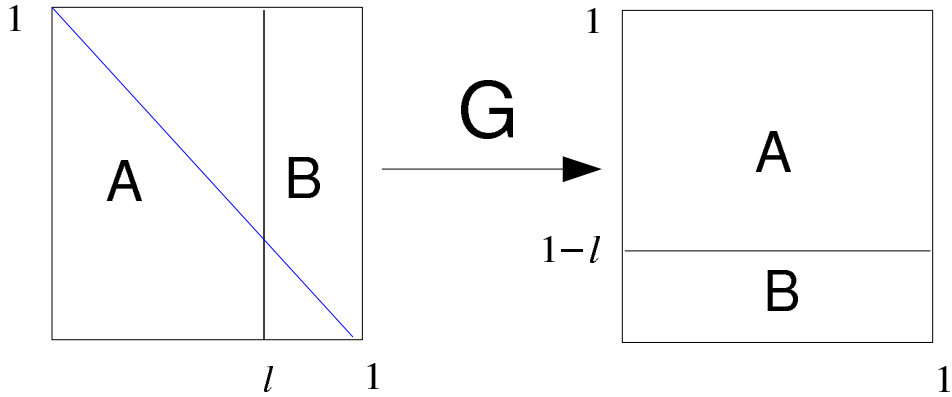}\\
  \caption{Involution $G$ for the map defined by
  Eq.(\ref{Map1}).}\label{involM1} \end{center}
\end{figure}

At each iteration, $\mathcal{U}$ is mapped onto itself,
and the Jacobian determinant is given by
\be
J_M(\un{x})\left\{
  \begin{array}{ll}
    J_A= r/l, & \hbox{for $\:0\leq x \leq l$;} \\
    J_B= l/r=J_A^{-1}, & \hbox{for $\:l\leq x \leq 1$.}\label{JDorf}
  \end{array}\right. \quad .
\ee
The map $M$ is locally either phase space contracting or expanding.
Furthermore, the constraint $r+l=1$ makes the map reversible, in the
sense of admitting the following involution $G$, meant to mimic the
time reversal invariant nature of the equations of motion of a
particle system,
\be
\left(\begin{array}{c}
  x_G \\
  y_G
\end{array}\right)=G \left(\begin{array}{c}
  x \\
  y
\end{array}\right)\left(\begin{array}{c}
  1-y \\
  1-x
  \end{array}\right) \quad .
\ee
The map $G$ amounts to a simple mirror symmetry operation with respect to
the diagonal represented in Fig.\ref{involM1}.

\begin{figure}[tbh]
\begin{center}
  \includegraphics[width=12cm, width=\textwidth]{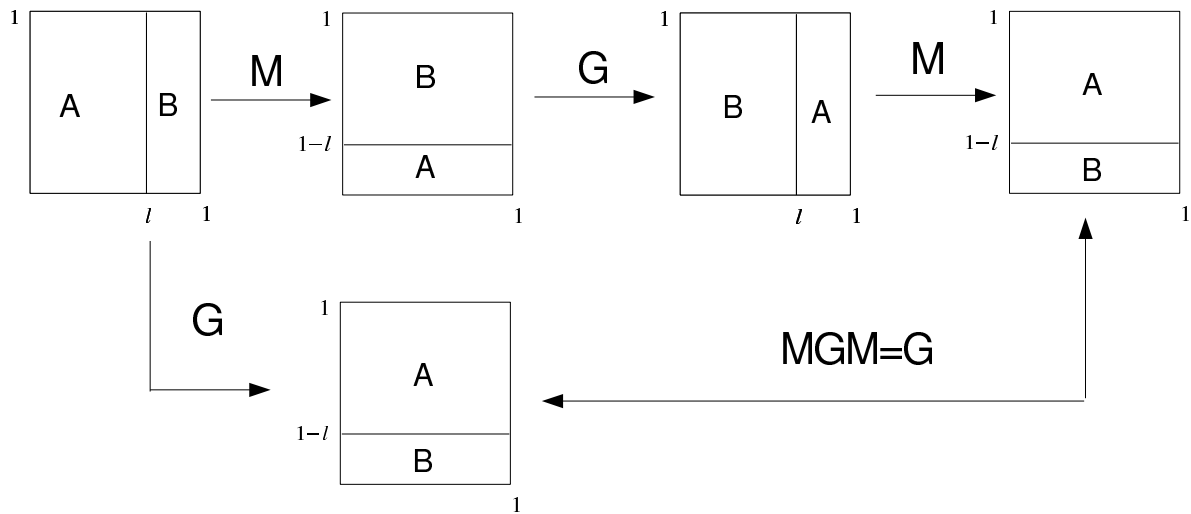}\\
  \caption{Check of reversibility for the map in Eq.(\ref{Map1}),
  performed by verifying Eq.(\ref{reversible}).}\label{checkinvolM1}
  \end{center}
\end{figure}

The relation $J_B=J_A^{-1}$ in Eq.(\ref{JDorf}) is a direct
consequence of the time reversibility of the model. To see how this
occurs, let us first observe, with the aid of Fig. \ref{checkinvolM1},
that the following relations hold for the map Eq.(\ref{Map1}):
\be
\hspace{3cm} G M A = B ~, \quad G M B  = A \quad .
\label{GM1}
\ee
Combining this with Eq.(\ref{general}), we immediately obtain
Eq.(\ref{JDorf}). Relation (\ref{general}) can be further exploited by
introducing the Jacobians of the dynamics restricted to the stable and
unstable manifolds in the generic regions $i = \{A,B\}$, which we
denote by $J_i^{s}$ and $J_i^{u}$, respectively. One then has
$$
J^{u}_{A} J^{s}_{A}=(J^{u}_{B})^{-1}(J^{s}_{B})^{-1}
$$
which, considering the specific constraints of our map,
$(J^{u}_{A})^{-1}+(J^{u}_{B})^{-1}=1$ and $J^{s}_{A}+J^{s}_{B}=1$,
leads to
\be
J^{s}_{A}=(J^{u}_{B})^{-1} \quad \mbox{ and } ~
J^{s}_{B}=(J^{u}_{A})^{-1} \quad .
\label{RecSU}
\ee
These equations constitute a consequence,
like Eq.(\ref{general}), of the time-reversibility of the model
\cite{Ruelle}.

A probability density $\rho_n$ on $\mathcal{U}$, given at time $n$,
evolves according to the Frobenius-Perron equation as \cite{Dorfman,Lasota}
\be
\varrho_{n+1}(M(\un{x}))=J_M^{-1}(\un{x})\: \varrho_{n}(\un{x})
\quad . \label{Liouville}
\ee
Correspondingly, the mean values of a phase function $\mathcal{O} :
\mathcal{U} \to \zR$ evolve and can be computed as
\be
\langle \mathcal{O} \rangle_n = \int_\mathcal{U} \mathcal{O}(\un{x}) {\rm d} \mu_n(\un{x}) \int_\mathcal{U} \mathcal{O}(\un{x}) \rho_n(\un{x}) ~ {\rm d} \un{x} \quad .
\ee
If $\langle \mathcal{O} \rangle_n$ converges exponentially to a given
steady state value $\langle \mathcal{O}\rangle$, for all phase
variables $\mathcal{O}$,\footnote{The space of phase functions depends
on the purpose one has in mind. The choice of H\"older continuous
functions is common \cite{Ruelle}.} one says that the state
represented by the regular measure $\mu_n$ corresponding to the
density $\rho_n$ converges to a steady state, which yields the
asymptotic time statistics of the dynamics. This state will be
characterized by an invariant measure $\mu$, which typically is a
natural one. For our models this measure is singular, because $M$ is
dissipative \cite{TGD98,Dorfman}.

However, due to the definition of the map Eq.(\ref{Map1}), which
stretches distances in the horizontal direction $-$the direction of
the unstable manifolds$-$, every application of the map smoothes any
initial probability density in that direction, so that our invariant
measure is uniform along the $x$-axis. Therefore, to compute steady
state averages it is not necessary to use the full information
provided by the $n \to \infty$ limit of Eq.(\ref{Liouville}).  Without
loss of generality, we may assume that the initial state is
``microcanonical'', i.e.\ its density is uniform in $\mathcal{U}$,
$\rho_0(x,y) = 1$.  Then each iteration of the map keeps the density
uniform along $x$, while it produces discontinuities in the $y$
direction, so that the $n$-th iterate of the density can be factorized
as
\be
\rho_n(x,y) = C \cdot \hat{\rho}_n(y) \quad ,
\ee
where $\hat{\rho}_n$ is a piecewise constant function, which gradually
builds up to a fractal structure, and $C$ is a constant that is easily
computed to be $1$ by requiring the normalization of $\rho_n$.
Hence, the varying averages of observables are computed as
\be
\langle \mathcal{O} \rangle_n \int_0^1 {\rm d} x  \int_0^1 {\rm d} y ~ \mathcal{O}(x,y) \hat{\rho}_n(y) \quad ,
\ee
and their steady state values are obtained by taking the limit $n \to
\infty$.  The average of the phase space contraction rate,
which is constant along the $y$-axis, is then easily obtained as
\bea
\langle \zL \rangle_n &=& -\int_0^1 {\rm d} x  \int_0^1 {\rm d} y ~ \hat{\rho}_n(y) \ln J(x)  \nonumber\\
&=& \int_0^l {\rm d} x \ln \frac{r}{l} + \int_l^1 {\rm d} x \ln \frac{l}{r} (l-r) \ln (l/r) \label{invden1} \quad .
\eea
As this result does not depend on $n$, it does not change by taking
the limit, and we have $\langle \zL \rangle = (l-r) \ln (l/r)$, which
vanishes for $l=1/2$ and is positive for all other $l \in (0,1)$.
From Eqs.(\ref{sigmadir}) and (\ref{JDorf}), we can write
\be
n ~ \overline{\zL}_{n}=(\alpha-\beta)\ln J_B \label{sigmadorf} \quad ,
\ee
where $\alpha$ and $\beta=n-\alpha$ denote the number of times the
trajectory falls in region $A$ or region $B$, respectively.

To proceed with the derivation of the $\zL$-FR for this map, one may
now follow two equivalent approaches. First of all, observe that our
map is of Anosov type, except for an inessential line of
discontinuity, which does not prevent the existence of a Markov
partition. Therefore, two basic approaches to the proof of the
$\zL$-FR may be considered: One may either trust the expansion of the
invariant measure in terms of unstable periodic orbits \cite{MR94,ChaosBook}, or one may adopt a stochastic approach to the
fluctuation relation \cite{Dorfman}, motivated by the fact that our
baker map is isomorphic to a Bernoulli shift, i.e.\ to a Markov chain
whose transition probabilities fulfill
$$
p(i_{M^k
(\un{x}_0)};k \rightarrow i_{M^{k+1} (\un{x}_0)};k+1) = p(i_{M^{k+1}
(\un{x}_0)};k+1) \quad ,
$$
where $i_{M^k (\un{x}_0)}$, with $k\in[0,n-1]$, denotes the region
containing the point $M^k (\un{x}_0)$, out of the two regions $\{ A,
B\}$, $p(i_{M^k (\un{x}_0)};k \rightarrow i_{M^{k+1} (\un{x}_0)};k+1)$
denotes the probability that the evolution touches region
$i_{M^{k+1}(\un{x}_0)}$ at the time step $k+1$, given that it visited
the region $i_{M^k (\un{x}_0)}$ at the previous time step $k$, and
$p(i_{M^{k+1} (\un{x}_0)};k+1)$ is the probability that
$M^{k+1}(\un{x}_0)$ belongs to the region $i_{M^{k+1} (\un{x}_0)}$. In
the $n
\rightarrow \infty$ limit, the latter becomes the invariant measure
$\mu_{i_{M^{k+1} (\un{x}_0)}}$ of the region $i_{M^{k+1} (\un{x}_0)}$
itself.

If one uses unstable periodic orbits, the argument proceeds as
follows: Every orbit $\zw$ is assigned a weight proportional to the
inverse of the Jacobian determinant of the dynamics restricted to its
unstable manifold, which is $J_\zw^{u}=(J^{u}_A)^{\za}
(J^{u}_B)^{\zb}$, if $\zw$ falls in region $A$ a number $\za$ of times
and falls in region $B$ a number $\zb$ of times.  Then the probability
that the dimensionless phase space contraction rate $e_{n}$, computed
over a segment of a typical trajectory, falls in the interval
$B_{p,\zd}=(p-\zd,p+\zd)$, coincides, in the large $n$ limit, with the
sum of the weights of the periodic orbits whose mean phase space
contraction rate falls in $B_{p,\zd}$. Denoting this steady state
probability by $\pi_n(B_{p,\zd})$, one can write
\begin{equation}
\pi_n(B_{p,\zd}) \approx \frac{1}{N_n} ~
 \sum_{\zw,e_{n}(\zw)\in B_{p,\zd}} (J_\zw^{u})^{-1} \quad ,
\label{Piofp}
\end{equation}
where $N_n$ is a normalization constant, and the approximate equality
becomes exact when $n \to \infty$. Because the support of the
invariant measure is the whole phase space $\mathcal{U}$, time reversibility
guarantees that the support of $\pi_n$ is
symmetric around $0$, and one can consider the ratio
\begin{equation}
\frac{\pi_n(B_{p,\zd})}{\pi_n(B_{-p,\zd})} \approx
\frac{\sum_{\zw,e_{n}(\zw) \in B_{p,\zd}} (J_\zw^{u})^{-1}}
{\sum_{\zw,e_{n}(\zw)\in B_{-p,\zd}} (J_\zw^{u})^{-1}} ~,
\label{pminusp}
\end{equation}
where each $\zw$ in the numerator has a counterpart in the
denominator, and the two are related through the involution $G$, as
implied by Eq.(\ref{property}). Therefore, considering each pair of
trajectory segments $\zw$ and $\overline{\zw}$, of initial conditions
$\un{x}_0$ and $G M^{n}(\un{x}_0)$ respectively, Eqs.(\ref{property})
and (\ref{RecSU}) imply
\begin{equation}
e_{n}(\zw)=-\overline{e}_{n}(\overline{\zw}) ~, \quad
(J_{\overline{\zw}}^{u})^{-1} = J_\zw^{s} ~,
\end{equation}
where, for sake of simplicity, by $e_{n}(\zw)$ we mean the average of
$e_n(\un{x}_0)$, based on any point $\un{x}_0$ of the orbit $\zw$.
Consequently, exponentiating the definition of $e_n \overline{\zL}_n/\langle \zL \rangle$, and recalling that $J_\zw J_\zw^s J_\zw^u$, for every orbit $\zw$, we may write
\be
\frac{J_{\overline{\zw}}^{u}}{J_{\zw}^{u}} \frac{1}{J_{\zw}^{s} J_{\zw}^{u}} = \frac{1}{J_{\zw}} \exp\left[ n \left( \langle \zL \rangle p + \ze_\zw \right) \right]
\label{Cpunter}
\ee
where $|\ze_\zw| \le \zd$ if $e_n(\zw)
\in B_{p,\zd}$.  Because each forward orbit $\zw$ in the denominator of
Eq.(\ref{pminusp}) has a counterpart $\overline{\zw}$ in the
denominator, and Eq.(\ref{Cpunter}) holds for each such pair, apart
from an error bounded by $\zd$, the whole expression Eq.(\ref{pminusp})
takes the same value as each of the ratios Eq.(\ref{Cpunter}), with an
error $|\ze| \le \zd$,
\begin{equation}
\frac{\pi_{n}(B_{p,\zd})}{\pi_n(B_{-p,\zd})}
= e^{n \left( \langle \zL \rangle p + \ze \right)} ~,
\label{largedev}
\end{equation}
where $\ze$ can be made arbitrarily small by taking $\zd$ sufficiently
small and $n$ sufficiently large. For a given $\zd$, $n$ must also be
large because, at every finite $n$, the values which $e_n$ takes
constitute $2n+1$ isolated points in $[-1,1]$. Therefore,
$\pi_{n}(B_{p,\zd})$ vanishes if none of these values falls in
$B_{p,\zd}$, making the expression senseless. But the set of these
values becomes denser and denser as $n$ increases. Taking the
logarithm of Eq.(\ref{largedev}), for consistency with
Eq.(\ref{prethm}), and choosing $p$ among the values $e_n$ which may
be attained along a periodic orbit of period $n$, we may now write
\be
\frac{1}{n \langle \zL \rangle} \ln\frac{\pi_{n}(B_{e_n,\zd})}{\pi_{n}(B_{-e_n,\zd})} = e_n \frac{1}{n \langle \zL \rangle} (\alpha-\beta)\ln\left(\frac{l}{r}\right)
\ee
for any $\zd > 0$. The $n \to \infty$ limit of the above expressions
confirms the validity of the $\zL$-FR, under the assumption that the
unstable periodic orbit expansion could be applied.

From the point of view of the Bernoulli shift description we obtain the
same result, supporting the applicability of the the unstable periodic
orbit expansion, despite the discontinuity of the dynamics
at $l$. Indeed, observe that $l$ equals the probability
$\mu_A=\int_A \mu({\rm d} \un{x})$ that the trajectory can be found
in region $A$, and $r$ equals the probability $\mu_B=\int_B \mu({\rm d} \un{x})$
that it is found in region $B$. Therefore, one may write as well
\be
\ln\frac{\pi_{n}(B_{e_n,\zd})}{\pi_{n}(B_{-e_n,\zd})} \ln\frac{\mu_{A}^{\alpha}\mu_{B}^{\beta}}{\mu_{A}^{\beta}\mu_{B}^{\alpha}}
\label{ratio1} \quad ,
\ee
which is due to the instantaneous decay of correlations in the
Bernoulli process.  This leads us to conclude that the violation of
the Anosov property, in this simple baker model, is irrelevant for its
behavior.

\section{The $\Lambda$-FR for a generalized dissipative baker map}
\label{sec:sec4}

\begin{figure}
\begin{center}
  \includegraphics[width=12cm, width=0.90\textwidth]{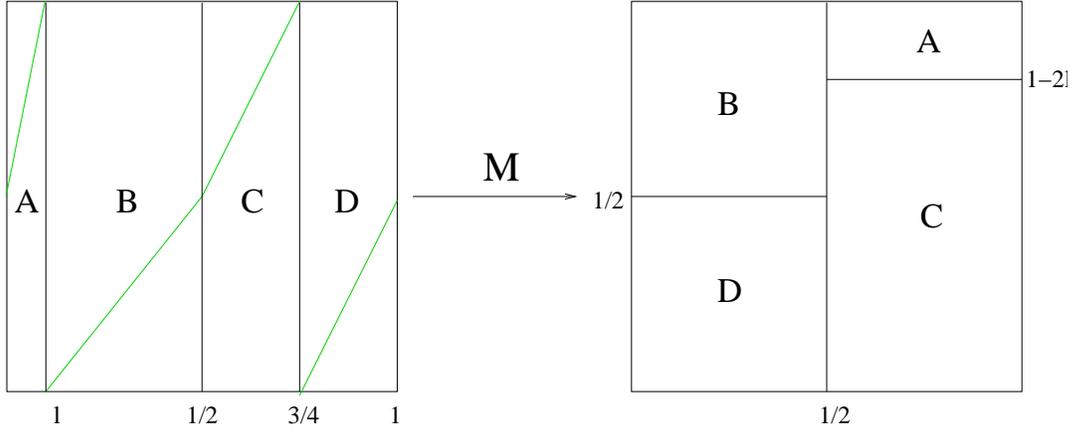}\\
  \caption{Illustration of the generalized baker map.  \textit{Green
  lines}: Piecewise linear one-dimensional map, which generates the
  dynamics along the unstable manifold.}\label{M2}
\end{center}
\end{figure}

We now propose a novel, generalized baker map, which is different from
previous models
\cite{Dorfman,VTB,GiDo99,TGD98,Voll02} by generating a discontinuity
in the invariant density along the $x$-axis. As illustrated in
Fig.~\ref{M2}, this is achieved by the map acting differently on four
subregions of $\mathcal{U}=[0,1]\times[0,1]$, defined by
\be
\left(\begin{array}{c}
  x_{n+1} \\
  y_{n+1}
\end{array}\right) = M
\left(\begin{array}{c}
  x_{n} \\
  y_{n}
  \end{array}\right)=\left\{
  \begin{array}{ll}
        \left(\begin{array}{c}
            \frac{1}{2l}x_{n}+\frac{1}{2} \\ \\
            2 l y_{n} +1-2l
          \end{array}\right), & \hbox{for $0\leq x < l$;} \\
            \left(\begin{array}{c}
            \frac{1}{1-2l}x_{n}-\frac{l}{1-2l}\\
            \frac{1}{2}y_{n} +\frac{1}{2}
          \end{array}\right), & \hbox{for $l\leq x < \frac{1}{2}$;} \\
            \left(\begin{array}{c}
            2x_{n}-\frac{1}{2} \\
            (1-2 l) y_{n}
          \end{array}\right), & \hbox{for $\frac{1}{2}\leq x < \frac{3}{4}$;} \\
            \left(\begin{array}{c}
            2x_{n}-\frac{3}{2} \\
            \frac{1}{2} y_{n}
          \end{array}\right), & \hbox{for $\frac{3}{4}\leq x \leq 1$.}
  \end{array}
\right. \label{Map2} \: .
\ee

\begin{figure}[htb]
\begin{center}
  \includegraphics[width=\textwidth]{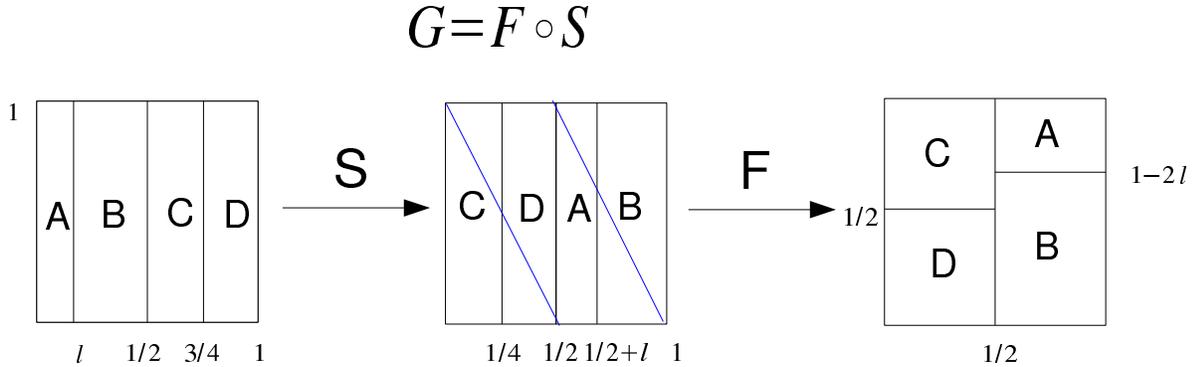}\\
  \caption{Involution $G$ for the map defined by
  Eq.(\ref{Map2}).}\label{involM2} \end{center}
\end{figure}

In the sequel, unless stated otherwise, by $M$ we refer to the map
introduced in Eq.(\ref{Map2}).  The model is fixed by choosing the
value of $l\in[0,\frac{1}{4}]$, i.e.\ the width of sub-region $A$.
The parameter which determines the dissipation, hence the
nonequilibrium steady state, corresponds to a bias $b$ which is
suitably defined by $b=J_C^u-J_B^u=2-\frac{1}{1-2l}$.  According to
its geometric construction shown in Fig.~\ref{M2}, the map
Eq.(\ref{Map2}) is area-contracting in region $B$, area-expanding in
region $C$, and area preserving in regions $A$ and $D$. This is
confirmed by computing the local Jacobian determinants of the map to
\be
J_{M}(\un{x})=\left\{
  \begin{array}{ll}
    J_A= 1, & \hbox{for \quad $0\leq x < l$;} \\
    J_B= [2(1-2l)]^{-1}, & \hbox{for \quad$l\leq x < \frac{1}{2}$;} \\
    J_C= 2(1-2l), & \hbox{for \quad$\frac{1}{2}\leq x < \frac{3}{4}$;} \\
    J_D= 1, & \hbox{for \quad $\frac{3}{4}\leq x \leq 1$.}\label{J2}
  \end{array}\right. \; .
\ee
The following involution $G$,
\be
\left(\begin{array}{c}
  x_{G} \\
  y_{G}
\end{array}\right)
  = G
\left(\begin{array}{c}
  x \\
  y
  \end{array}\right)\left\{
  \begin{array}{ll}
     \left(\begin{array}{c}
           \frac{1}{2}-\frac{y}{2} \\
            1-2x
          \end{array}\right), & \hbox{for \quad $0\leq x < \frac{1}{2}$;} \\
    \left(\begin{array}{c}
            1-\frac{y}{2} \\
            2-2x
          \end{array}\right), & \hbox{for \quad $\frac{1}{2}\leq x \leq 1$.}
  \end{array}
\right.\label{G}\: ,
\ee
constitutes a time reversal operator for the map $M$ defined on the
unit cell. It consists of the composition $G=F
\circ S$ of two other involutions, with $S$ permuting the left and the right
halfs of the unit square, and $F$ mirroring the regions along their
respective diagonals for all values $b \in (-\infty,1]$, cf.\ Fig.\
\ref{involM2}.

\begin{figure}[htb]
\begin{center}
  \includegraphics[width=\textwidth]{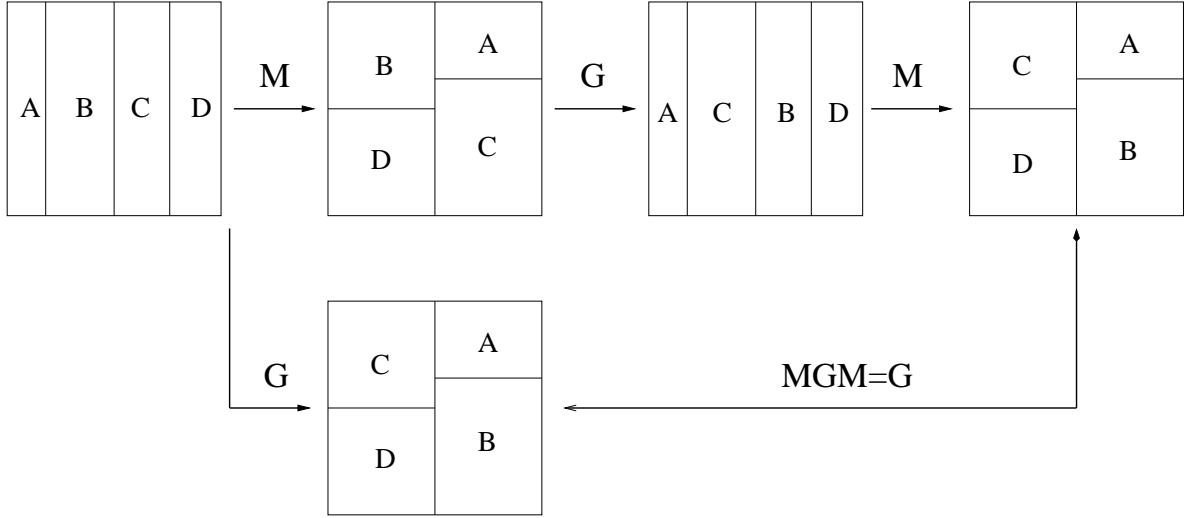}\\
  \caption{Check of reversibility for the map of Eq.(\ref{Map2}), performed by verifying Eq.(\ref{reversible}).}\label{checkinvolM2}
    \end{center}
\end{figure}

Analogously to Eq.(\ref{GM1}) for the map Eq.(\ref{Map1}), for the
generalized map Eq.(\ref{Map2}), Eq.(\ref{G}) entails the relations
\be
G M A = A ~, \quad G M D = D ~, \quad G M B = C ~, \quad G M C = B \label{GM2} \quad ,
\ee
which can also be inferred graphically from Fig.~\ref{checkinvolM2}.
It is readily seen, again, that the Jacobian rule Eq.(\ref{general}),
supplemented by Eq.(\ref{GM2}), implies the relations Eq.(\ref{J2}).

\begin{figure}
\begin{center}
  \includegraphics[width=12cm]{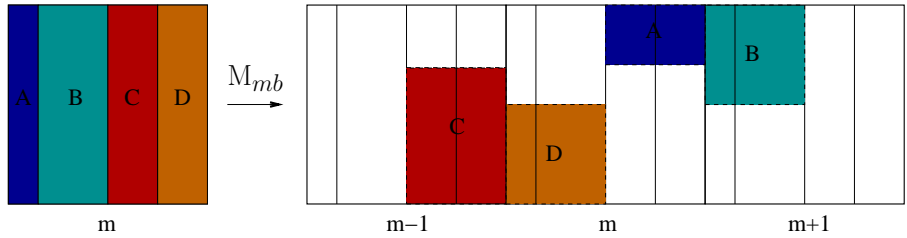}\\ \caption{Illustration of
  a multibaker chain based on the unit cell defined in
  Fig. \ref{M2}, featuring a flow of particles from the regions $B$
  and $C$ of the cell $m$ into, respectively, the neighboring cell
  $m+1$ on the right and onto the neighboring cell $m-1$ on the
  left. The net flow of particles corresponds to the current $\Psi$,
  which is found to be proportional to the average phase space
  contraction rate $\langle \zL \rangle$.}\label{multibaker}
\end{center}
\end{figure}

Let us now lift this biased dissipative baker map onto the whole real
line in form of a so-called multibaker map, which consists of an
infinitely long chain of baker unit cells deterministically coupled
with each other. Multibakers have been studied extensively over the
past two decades as simple models of chaotic transport
\cite{PGbook,RKbook,Voll02,VTB,RCmultibak,GiDo99}.
In our model, which we denote by $M_{mb}$, all unit cells are coupled
by shifting the regions $B$ and $C$ to the, respectively, right and
left neighboring cells, cf.\ Fig. \ref{multibaker}. Choosing $b\neq
0$, i.e.\ $l\neq\frac{1}{4}$ then implies the existence of a current
$\Psi(b)$, defined by the net flow of points from cell to cell.  The
map $M_{mb}$ is area contracting (expanding) in the direction
(opposite to the direction) of the current, analogously to the case of
typical thermostatted particle systems
\cite{VTB,RKbook}.\footnote{Differently, the pump model of
Ref.\cite{BenRonPump} may be tuned to expand phase space volumes in
the direction of the current.} This can be inferred from the graphical
construction in Fig. \ref{multibaker} complemented by the relations
Eq.(\ref{J2}).

To asses the validity of the $\zL$-FR for this model, let us observe
that the form of the invariant probability distribution along the
$y$-direction (the direction of the stable manifolds) is irrelevant,
analogously to the case discussed in Sec.\ref{sec:sec3}, because the
phase space contraction per time step, $\zL$, does not depend on
$y$. By introducing the shorthand notation $\phi=\ln J_C$ we have
\be
\Lambda(\un{x}) = \Lambda(x,y) \left\{
  \begin{array}{ll}
    0, & \hbox{for $0\leq x < l$;} \\
    \phi, & \hbox{for $l\leq x < \frac{1}{2}$;} \\
    -\phi, & \hbox{for $\frac{1}{2}\leq x < \frac{3}{4}$;} \\
    0, & \hbox{for $\frac{3}{4}\leq x \leq 1$.}
  \end{array}
\right.\label{1}\quad .
\ee
The $y$-coordinate may then be integrated out, and one only needs to
consider the projection of the invariant measure on the $x$ axis, the
direction of the unstable manifolds, which has density $\rho_x$.

The calculation of this invariant density can be conveniently
performed by introducing a Markov partition of the unit interval,
which separates the region $0\leq x < 1/2$ from the region $1/2\leq x
\leq 1$. Denote by $\rho_{l}$ and $\rho_{r}$ the projected density
computed in these two regions and let $T$ be the transfer operator
associated with the Markov partition. One may then compute the
evolution of the projected densities, which are now piecewise
constant, if the initial distribution is uniform on the unit
square. In this case the corresponding Frobenius-Perron equation
Eq.(\ref{Liouville}) takes the form \cite{PGbook,GaKl98,RKbook}
\be
\left(\begin{array}{c}
        \rho_{l}(x_{n+1}) \\
        \rho_{r}(x_{n+1})
      \end{array}\right)=T\cdot \left(\begin{array}{c}
        \rho_{l}(x_{n}) \\
        \rho_{r}(x_{n})
      \end{array}\right)
~, \qquad
T=\left(
    \begin{array}{cc}
      1-2l & 1/2 \\
      2l & 1/2 \\
    \end{array}
  \right) \quad . \label{toptm}
\ee
According to the Perron-Frobenius theorem, the transfer matrix $T$ has
largest eigenvalue $\lambda=1$, whose corresponding eigenvector yields
the invariant density of the system to
\be
\rho(x)\left\{
  \begin{array}{ll}
    \rho_l(x)=\frac{2}{1+4l}, & \hbox{for \quad $0\leq x < \frac{1}{2}$;} \\
    \rho_r(x)=\frac{8l}{1+4l}, & \hbox{for \quad $\frac{1}{2}\leq x \leq 1$.}
  \end{array}
\right. \; . \label{SRB2}
\ee
This result confirms that, by construction of the model, and
differently from the one considered in Section \ref{sec:sec3}, the
density of the map Eq.(\ref{Map2}) is not uniform along the
$x$-direction, that is, it is actually discontinuous along the
unstable direction.

By using this density, the average phase space contraction rate can be
calculated to
\be
\langle \zL \rangle =-\Psi(b)\ln\frac{2-b}{2}\ge0 \quad ,\label{oursigma}
\ee
where
\be
\Psi(b)=\frac{b}{4-3b}
\ee
is the steady state current in the corresponding multibaker chain.
Note that
\be
\Psi(b)\to\frac{b}{4}\quad(b\to0) \quad ,
\ee
hence we have linear response and a caricature of Ohm's
law. Accordingly, we get
\be
\langle \zL \rangle \to\frac{b^2}{8}\quad(b\to0)
\ee
for the average phase space contraction rate, as one would expect from
nonequilibrium thermodynamics if this quantity was identified with the
nonequilibrium entropy production rate of a system
\cite{VTB,Voll02}. This confirms that our abstract map represents a
`reasonably good toy model' in capturing some properties as they are
expected to hold for ordinary nonequilibrium processes. Related biased
one-dimensional maps have been studied in
Refs.~\cite{GrKl02,RKbook}. Note that $\Psi=0$ for $l=1/4$,
respectively $b=0$, only, in which case the dynamics is conservative,
and the model boils down to a special case of the multibaker map
analyzed in Ref.~\cite{GaKl98}.


In order to check the $\Lambda$-FR for this model, we first
need to define the transition probabilities $p_{ij}$ of jumping from
region $i$ to region $j$, with $i,j \in \{A,B,C,D\}$ denoting the
finite state space. They constitute the elements of the transition
matrix
\be
 P=\left(
     \begin{array}{cccc}
       0 & 0 & \frac{1}{2} & \frac{1}{2} \\
       2l & 1-2l & 0 & 0 \\
       0 & 0 & \frac{1}{2} & \frac{1}{2} \\
       2l & 1-2l & 0 & 0 \\
     \end{array}
   \right) \label{P} \quad .
 \ee

Note that $P$ defines a {\em stochastic} transition matrix, which acts
onto vectors whose elements are the probabilities to be in the
different regions, in contrast to the {\em topological} transition
matrix Eq.(\ref{toptm}), which acts upon probability density vectors.
The left eigenvector of $P$, associated with the
eigenvalue $1$, corresponds to the vector of the invariant
probabilities $\mu_i$ of the regions $A, B, C$ and $D$.
Alternatively, since the projected invariant probability density is
constant in each of these four regions, the $\mu_i$'s are also
immediately obtained by multiplying the relevant invariant density
Eq.(\ref{SRB2}) with the width of the respective region. One way or the
other, we obtain
\be
\mu_i\left\{
  \begin{array}{ll}
    \frac{2l}{1+4l}, & \hbox{if ~ $i=A,C,D$;} \\
    \frac{1-2l}{1+4l}, & \hbox{if ~ $i=B$.}
  \end{array}
\right.\label{mu}\quad .
\ee
The discontinuity of the invariant density Eq.(\ref{SRB2}) along the
unstable direction, for $l \ne 1/4$, means that the Anosov property is
more substantially violated here than for the map in Section
\ref{sec:sec3}.
Therefore, the periodic orbit expansion used in Section \ref{sec:sec3}
cannot be immediately trusted, and an alternative method is better
suited to prove the validity of the $\zL$-FR.

We may begin by considering a trajectory segment of $n$ steps, which
starts at $\un{x}_0 \in i_{\un{x}_0}$ and ends in $\un{x}_n \in
i_{\un{x}_n}$, hence visits the regions
$\{i_{\un{x}_0},...,i_{\un{x}_n}\}$. Consider the first $(n-1)$
transitions, corresponding to the symbol sequence
$\{i_{\un{x}_0},...,i_{\un{x}_{n-1}}\}$, and treat separately the last
transition $i_{\un{x}_{n-1}}\rightarrow i_{\un{x}_n}$. Denote by
$n_{ij}$ the number of transitions from region $i$ to region $j$,
along the trajectory segment of $(n-1)$ steps, and by $n_{i}=\sum_{\{
j: p_{ij}\neq 0\}} n_{ij}$ the total number of transitions starting in
$i$.  Some transitions are forbidden, as shown by Eq.(\ref{P}), hence
the following holds:
\be
\underbrace{n_{AC}+n_{AD}}_{n_A}+\underbrace{n_{BA}+n_{BB}}_{n_B}+\underbrace{n_{CC}+n_{CD}}_{n_C}+\underbrace{n_{DA}+n_{DB}}_{n_D}=n-1\nonumber \quad .
\ee
It also proves convenient to introduce the following symbols:
 \be
 n_{\rightsquigarrow i}\left\{
  \begin{array}{ll}
    0, & \hbox{if the trajectory does not start in $i$;} \\
    1, & \hbox{if the trajectory starts in $i$.}
  \end{array}
\right.
\ee
\be
 n_{i \rightsquigarrow}\left\{
  \begin{array}{ll}
    0, & \hbox{if the trajectory does not end in $i$;} \\
    1, & \hbox{if the trajectory ends in $i$.}
  \end{array}
\right.
\ee
and $\Delta_{ij}= n_{\rightsquigarrow i}-n_{j \rightsquigarrow}$. The
quantities $n_{\rightsquigarrow i}$ and $n_{i \rightsquigarrow}$ take
into account the possibility that the trajectory segment may,
respectively, start in, or end into, the region $i$.  Thus, we may
write the following \textit{flux balances}
\be
\sum_{\{ i: p_{ij}\neq 0\}} n_{ij}=n_{j}-\Delta_{jj}, \quad \forall j \label{flux}
\ee
for each region of the map.
Next, we introduce the quantity
\be
g=n_B-n_C+n_{B \rightsquigarrow}-n_{C \rightsquigarrow} \label{gg}\quad ,
\ee
which lies in the interval $[-n,n]$ and is related by
\be
\overline{\zL}_{n}={g} \phi/n \label{pscg}
\ee
to the average phase space contraction in a trajectory segment of $n$
steps.

To evaluate the ratio of probabilities appearing in the $\Lambda$-FR,
let us denote by $i_{\un{x}}$ the region containing the point
$\un{x}$, out of the four regions $\{ A, B, C, D \}$, and let us focus
on a single trajectory of initial condition $\un{x}_0=(x_0,y_0) \in
i_{\un{x}_0}$.  For a given $n$, the sequence of transitions which
take this point from region $i_{\un{x}_0}$ to region
$i_{M(\un{x}_0)}=i_{\un{x}_1}$, from region $i_{\un{x}_1}$ to region
$i_{M^2(\un{x}_0)}=i_{\un{x}_2}$ and eventually from region
$i_{M^{n-1}(\un{x}_0)}=i_{\un{x}_{n-1}}$ to region
$i_{M^n(\un{x}_0)}=i_{\un{x}_{n}}$ does not depend on $y_0$. The
larger the value of $n$, the narrower the width of the set of initial
conditions whose trajectories undergo the same sequence of $n$
transitions experienced by the trajectory starting in $\un{x}_0$. Let
$\omega(\un{x}_0,n) = \{\un{x} \in \mathcal{U}: M^k (\un{x}) \in
i_{M^k(\un{x}_0)} , k = 0,...,n \}
\subset i_{\un{x}_0}$ denote this set of initial conditions. The expansiveness of the map implies
$$
\lim_{n \to \infty} \omega(\un{x}_0,n) = \left\{ \un{x}=(x,y) : x=x_0 , y \in [0,1] \right\} ~.
$$ Because the phase space contraction $\zL(\un{x}_k)$ only depends on
the region $i_{\un{x}_k}$ from which the transition $i_{\un{x}_k} \to
i_{\un{x}_{k+1}}$ occurs, all trajectory segments of $n$ steps
originating in $\omega(\un{x}_0,n)$ enjoy the same average phase space
contraction $\overline{\zL}_n$. The amount $\overline{\zL}_n$ is also
produced by the trajectory segments which visit the regions
$i_{\un{x}_0}, ..., i_{\un{x}_{n-1}}$ and eventually land in
$i_{\un{\hat{x}}_n}$, where $i_{\un{\hat{x}}_n} \ne i_{\un{x}_{n}}$ is
the other region reachable from $i_{\un{x}_{n-1}}$. Let
$\omega(\un{\hat{x}}_0,n)$ be this second set of initial conditions
producing $\overline{\zL}_n$ in $n$ steps. The point $\un{\hat{x}}_0$
lies in $i_{\un{x}_0}$, i.e.\ $i_{\un{\hat{x}}_0}=i_{\un{x}_0}$, but
differs from ${\un{x}_0}$ and does not belong to $\omega(\un{x}_0,n)$.
Denoting by $\pi_{\omega(\un{x}_0,n)}$ the invariant measure of
$\omega(\un{x}_0,n)$, one finds:
\bea
&&\pi_{\omega(\un{x}_0,n)} \nonumber \\
&&=\mu_{i_{\un{x}_0}} \prod_{k=0}^{n-2} p(i_{M^{k}\un{x}_0};k\rightarrow i_{M^{k+1}\un{x}_0};k+1)
p(i_{M^{n-1}(\un{x}_0)}, n-1\rightarrow i_{M^n(\un{x}_0)};n) \nonumber\\
\nonumber\\
&&= \mu_{i_{\un{x}_0}} p_{AC}^{n_{AC}}p_{AD}^{n_{AD}}p_{BA}^{n_{BA}}p_{BB}^{n_{BB}}p_{CC}^{n_{CC}}
p_{CD}^{n_{CD}}p_{DA}^{n_{DA}}p_{DB}^{n_{DB}} p(i_{M^{n-1}(\un{x}_0)}; n-1\rightarrow i_{M^n(\un{x}_0)};n)\nonumber\\
\nonumber\\
&&= \mu_{i_{\un{x}_0}}
p_{DA}^{n_{BA}+n_{DA}}p_{BB}^{n_{BB}+n_{DB}}p_{CC}^{n_{AC}+n_{CC}}
p_{AD}^{n_{AD}+n_{CD}} p(i_{M^{n-1}(\un{x}_0)}; n-1\rightarrow i_{M^n(\un{x}_0)};n)\nonumber\\
& \nonumber\\
&&=\mu_{i_{\un{x}_0}} p_{DA}^{n_A-\Delta_{AA}}p_{BB}^{n_B-\Delta_{BB}}p_{CC}^{n_C-\Delta_{CC}}
p_{AD}^{n_D-\Delta_{DD}} \times \nonumber\\
\nonumber\\
&& \times p(i_{M^{n-1}(\un{x}_0)}; n-1\rightarrow i_{M^n(\un{x}_0)};n)
\label{forw}\: ,
\eea
where we made use of Eqs.(\ref{flux}) and of the equalities
$p_{ij}=p_{kj}$ for all $i \neq j$, which can be deduced from an
inspection of Eq.(\ref{P}).  Similarly, one has
\be
\pi_{\omega(\un{\hat{x}}_0,n)} = \mu_{i_{\un{x}_0}}
p_{DA}^{n_A-\Delta_{AA}}p_{BB}^{n_B-\Delta_{BB}}p_{CC}^{n_C-\Delta_{CC}}
p_{AD}^{n_D-\Delta_{DD}} p(i_{M^{n-1}(\un{x}_0)}; n-1\rightarrow i_{\un{\hat{x}}_n};n) ~.
\label{forhat}
\ee
Given the similarity of the expressions Eqs.(\ref{forw}) and
(\ref{forhat}), and the fact that
\be
p(i_{M^{n-1}(\un{x}_0)}, n-1\rightarrow i_{M^n(\un{x}_0)};n) + p(i_{M^{n-1}(\un{x}_0)}; n-1\rightarrow i_{\un{\hat{x}}_n};n) = 1 ~,
\ee
it is convenient to consider the set
\be
\omega(\un{x}_0,n) \cup \omega(\un{\hat{x}}_0,n) = \omega(\un{x}_0,n-1)
\ee
whose measure is given by
\be
\pi_{\omega(\un{x}_0,n-1)}
=\mu_{i_{\un{x}_0}} p_{DA}^{n_A-\Delta_{AA}}p_{BB}^{n_B-\Delta_{BB}}p_{CC}^{n_C-\Delta_{CC}}p_{AD}^{n_D-\Delta_{DD}}\quad .
\label{union}
\ee
This measure represents the contribution to the probability of
producing $\overline{\zL}_n$ in $n$ steps, given by the trajectory
segments whose initial conditions lie in $\omega(\un{x}_0,n-1)$.  The
steady state probability of $\overline{\zL}_n$ is then the sum of
contributions like Eq.(\ref{union}), for all remaining sets of
trajectories compatible with $\overline{\zL}_n$, characterized by
distinct sequences of $n-1$ transitions.

As we discussed at the end of Section 2, for any initial point
$\un{x}_0$ in the phase space that experiences a mean phase space
contraction $\overline{\zL}_n(\un{x}_0)$ in $n$ steps, the point
$\un{x}_{0R} = G M M^{n-1} (\un{x}_0) = G M^n (\un{x}_0)$ experiences
the opposite mean phase space contraction
$\overline{\zL}_n(\un{x}_{0R}) = -\overline{\zL}_n(\un{x}_0)$, cf.\
Eq.(\ref{property}). The trajectory segment of $n$ steps, starting at
$\un{x}_{0R}$, is thus the time reversal of the one starting at
$\un{x}_0$, and ${\omega}(G M^n (\un{x}_0),n)$ is the set of initial
conditions of the time reversals of the segments beginning in
$\omega(\un{x}_0,n)$.  The segments beginning in ${\omega}(G M^n
(\un{x}_0),n)$ visit the regions $i_{G M^{n} (\un{x}_0)}$, $i_{G
M^{n-1} (\un{x}_0)}$, ... $i_{G (\un{x}_0)}$, hence they produce the
average phase space contraction $-\overline{\zL}_n$ if the segments
beginning in $\omega(\un{x}_0,n)$ produce $\overline{\zL}_n$. In
analogy to Eq.(\ref{forw}) their steady state probability is given by:
\bea
\pi_{{\omega}(G M^{n} (\un{x}_0),n)}&=& \mu_{i_{G M^n (\un{x}_0)}} \prod_{k=0}^{n-2} p(i_{G M^{n-k}(\un{x}_0)}; k \rightarrow i_{G M^{n-k-1}(\un{x}_0)};k+1)\times \nonumber\\
&\times & p(i_{G M (\un{x}_0)}; n-1\rightarrow i_{G(\un{x}_0)};n)\: .
\eea

\begin{figure}
  \begin{center} \includegraphics[width=10cm, height=5cm]{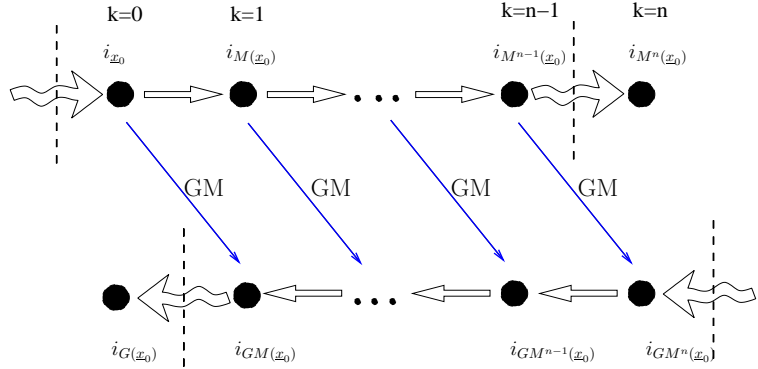}\\ \caption{Sequence of visited regions in the
  \textit{forward} (upper sequence) and \textit{time-reversed} (lower
  sequence) dynamics. The lower sequence is determined by applying the composite map $G\circ M$ onto the upper sequence, cf. Eq. (\ref{GM2}) (thick blue arrows).}\label{trajectory} \end{center}
\end{figure}

Again, this set of trajectories may be grouped together with the set
of trajectories whose last step falls in the other region reachable
from $G i_{\un{x}_1}$, say $\hat{i}_{\un{x}_0} \ne
i_{\un{x}_0}$. The probability of the union
${\omega}(G M^{n} (\un{x}_0),n-1)$ of these two sets takes the value
\bea
\pi_{\omega(G M^{n} (\un{x}_0),n-1)} &=&\mu_{i_{G M^n (\un{x}_0)}} \prod_{k=0}^{n-2} p(i_{G M^{n-k} (\un{x}_0)}; k \rightarrow i_{G M^{n-k-1} (\un{x}_0)};k+1) \nonumber \\
&=& \mu_{GM i_{M^{n-1} (\un{x}_0)}} p_{BA}^{n_{AC}}p_{DA}^{n_{AD}}p_{AC}^{n_{BA}}
p_{CC}^{n_{BB}}p_{BB}^{n_{CC}} p_{DB}^{n_{CD}}p_{AD}^{n_{DA}}p_{CD}^{n_{DB}} \nonumber\\
&=&\mu_{GM i_{M^{n-1} (\un{x}_0)}} p_{DA}^{n_A}p_{CC}^{n_B}p_{BB}^{n_C}p_{AD}^{n_D}
\label{rev}\quad ,
\eea
where we have made use, from Eq.(\ref{GM2}), of the crucial relation
$i_{GM^k (\un{x}_0)}=GMi_{M^{k-1} (\un{x}_0)}$, with $k=1,...,n$,
cf.\ Fig.~\ref{trajectory}.  This contribution to the probability of
producing $-\overline{\zL}_n$ in $n$ steps mirrors the contribution to
the probability of producing $\overline{\zL}_n$, given by
Eq.(\ref{union}). Taking the ratio of these two contributions and writing
the phase space contraction in terms of $g$ units of size $\phi$, cf.\
Eq.(\ref{pscg}), one obtains
\bea
\frac{\pi_{\omega(\un{x}_0,n-1)}}{\pi_{\omega(G M^{n} (\un{x}_0),n-1)}} &=&
\frac{\mu_{i_{\un{x}_0}}}{\mu_{GM i_{M^{n-1} (\un{x}_0)}}}
\frac{p_{DA}^{n_A-\Delta_{AA}}p_{BB}^{n_B-\Delta_{BB}}p_{CC}^{n_C-\Delta_{CC}}p_{AD}^{n_D-\Delta_{DD}}} {p_{DA}^{n_A}p_{CC}^{n_B}p_{BB}^{n_C}p_{AD}^{n_D}} \nonumber \\
 &=& \left(\frac{p_{BB}}{p_{CC}}\right)^g \alpha_\omega
\label{FR2}
\eea
with
\be
\alpha_{min} \leq \alpha_\omega=\left[\frac{\mu_{i_{\un{x}_0}}}{\mu_{GM i_{M^{n-1} (\un{x}_0)}}}p_{DA}^{-\Delta_{AA}}p_{BB}^{-\Delta_{BC}}p_{CC}^{-\Delta_{CB}}p_{AD}^{-\Delta_{DD}}\right]
\leq \alpha_{max} \quad ,
\label{alpha}
\ee
where the upper and lower bounds $\alpha_{min}$ and $\alpha_{max}$ are
$(g,n)$-independent positive numbers, which are found to be
\be
\alpha_{min}=4l=\alpha_{max}^{-1} \quad ,
\ee
as we numerically tested by considering all possible values of
$\alpha_\omega$ corresponding to a trajectory segment visiting the
regions $i_{\un{x}_0},...,i_{M^{n-1}(\un{x}_0)},i_{M^n(\un{x}_0)}$,
for any $i_{\un{x}_0},i_{M^{n-1}(\un{x}_0)}$ and $i_{M^n (\un{x}_0)}$.
At the same time, Eqs.(\ref{P}) and (\ref{mu}) imply the
equality being at the heart of the $\Lambda$-FR, i.e.\
$$
\left( \frac{p_{BB}}{p_{CC}}\right)^g = e^{g \phi}\quad .
$$ These results hold for all sets of trajectory segments starting in
${\omega}(\un{x}_0,n-1)$, related to their corresponding reversals
starting in ${\omega}(G M^{n} (\un{x}_0),n-1)$. Therefore,
Eq.(\ref{FR2}) holds as well for the total probabilities of producing
$\overline{\zL}_n$ and $-\overline{\zL}_n$, because the ratio of the
sums of the probabilities of the groups of trajectory segments
producing $\overline{\zL}_n$ and $-\overline{\zL}_n$ equals the ratio
of the probabilities of a single group, with corrections always bounded
by $\alpha_{min}$ and $\alpha_{max}$.

To match this result with the $\zL$-FR Eq.(\ref{prethm}), it now
suffices to introduce the normalized quantity $e_n = g \phi / n
\langle \Lambda \rangle$ and to take the logarithm of the ratio of
probabilities,
\be
e_n(\un{x}) - {\frac{\ln \alpha_{max}}{n \langle \Lambda \rangle}} \leq
\frac{1}{n \langle \Lambda \rangle} \ln\frac{\mu(\{ \un{x} : e_n(\un{x}) \in (p-\zd,p+\zd)\})}
{\mu(\{\un{x} : e_n(\un{x}) \in (-p-\zd,-p+\zd)\}}
\leq
e_n(\un{x}) +
{\frac{\ln \alpha_{max}}{n \langle \Lambda \rangle}} \: .
\ee
In the ${n\rightarrow\infty}$ limit, in which the allowed values of
$e_n$ become dense in the domain of the $\Lambda$-FR, one recovers the
fluctuation theorem with $p^*=\phi/\langle \zL \rangle$.

\section{Conclusions}
In this paper we have presented analytically tractable examples of
dynamical systems in order to clarify some aspects of the
applicability of the standard steady state fluctuation relation. In
our case, there is no distinction between the so-called $\zL$-FR and
$\zW$-FR, because the appropriate measure is the Lebesgue measure, in our case, cf.\
Eq.(\ref{omegat}) \cite{SearlRonEvans, RonMejia}. Our results show
that the $\zL$-FR holds under less stringent conditions than
those required by the Gallavotti-Cohen FT, which include time
reversibility and existence of an SRB measure, i.e. a measure which is
smooth along the unstable directions. This is of interest for
applications, because strong requirements such as the Anosov property
are hardly met by dynamics of physical interest, in general.

\begin{figure}
  \begin{center} \includegraphics[width=12cm,
  width=0.90\textwidth]{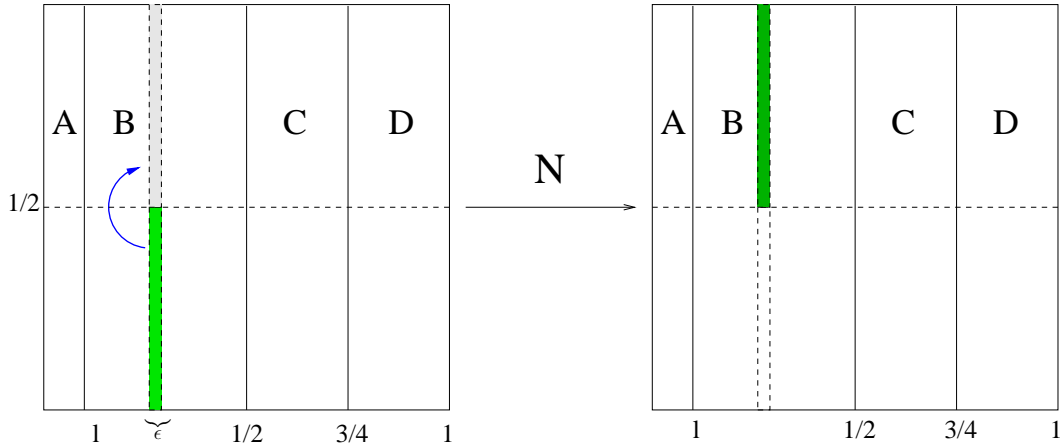}\\
\caption{The map $N$ defined in  Eq.(\ref{N}), which spoils the reversibility of the  model}\label{correction} \end{center}
\end{figure}

To obtain this result, we have considered an example in which the
involution representing the time reversal operator is discontinuous
\cite{Porta} and in which also the invariant measure is discontinuous
along the unstable direction. Our discontinuities are mild, as
discussed in the introduction, however, they illustrate how the
validity of the $\zL$-FR may be extended beyond the standard
constraints. Our proof capitalizes on the fact that the directions of
stable and unstable manifolds are fixed and that the vertical variable
does not affect the value of the phase space contraction rate. This
fact has rather profound implications concerning the validity of the
$\zL$-FR for cases in which time reversibility is more substantially
violated. In fact, only the knowledge of the forward and reversed
sequences of visited regions is required in order to verify the
$\zL$-FR, rather than the more detailed knowledge of the forward and
reversed trajectories in phase space.  Thus, for instance, one easily
realizes that our calculations may be carried out for a map of the
form $K = M \circ N$, where $M$ may refer to one of the maps Eqs.(\ref{Map1}) or
(\ref{Map2}), while $N$ does not contract or expand volumes and
affects in some irreversible fashion the $y$-coordinate only. $N$ can
be constructed in several ways: For example, let $M$ be the map
Eq.(\ref{Map2}), and assume that $N$ acts only on a vertical strip of
width $\epsilon$ in the region $B$, as follows:
\be
\left(\begin{array}{c}
  x_{n+1} \\
  y_{n+1}
\end{array}\right)
=N
\left(\begin{array}{c}
  x_{n} \\
  y_{n}
  \end{array}\right)
  \left\{
  \begin{array}{ll}
               \left(\begin{array}{c}
                   x_{n} \\
                   1- y_{n}
                 \end{array}\right), & \hbox{for $x \in [\tilde{x},\tilde{x}+\epsilon]$ and $y \in [0,\frac{1}{2}]$;} \\
    \left(\begin{array}{c}
                   x_{n} \\
                   y_{n}
                 \end{array}\right) & \hbox{for $x \in [\tilde{x},\tilde{x}+\epsilon]$ and $y \in (\frac{1}{2},1]$.}
  \end{array}
\right.\label{N} \: .
\ee
cf. Fig.\ref{correction} for a graphical representation.  The map $N$
is not reversible, according to the definition Eq.(\ref{reversible});
in fact, $N$ is not even a homeomorphism, as its inverse $N^{-1}$ is
not defined, so neither is the inverse of the composite map
$K^{-1}$. Nevertheless, the $\zL$-FR still holds in this case, due to
the existence of a milder notion of reversibility expressed by the
relations Eq.(\ref{GM2}). The latter entail that only a
\textit{coarse-grained} involution, mapping regions onto regions, is
needed for the proof of the $\Lambda$-FR, rather than a
\textit{local} involution, mapping points into points in phase space,
as defined by Eq.(\ref{reversible}).

\section{Acknowledgements}

M.C.\ acknowledges support from the Swiss National Science Foundation
(SNSF). R.K.\ is grateful to Ramses van Zon for helpful discussions on
time-reversibility. L.R.\ and P.D.G.\ gratefully acknowledge financial
support from the European Research Council under the European
Community's Seventh Framework Programme (FP7/2007-2013) / ERC grant
agreement n 202680.  The EC is not liable for any use that can be made
on the information contained herein.

\end{document}